# Magnetic Flux Trapping in Granular HTSC near Superconducting Transition

A. A. Sukhanov, V. I. Omelchenko, G. A. Orlova

Institute of Radioengineering and Electronics RAS, 141190, Vvedenskogo sq. 1, Fryazino, Moscow dist., Russia

e-mail: sukh@ms.ire.rssi.ru

**Abstract**

The temperature and field dependences of the trapped magnetic fields and of the frozen magnetoresistance of (Pb)Bi-Sr-Ca-Cu-O ceramics and Bi-based magnetron films are investigated. It is found that in the resistive transition region of granular Bi-HTSC the trapped magnetic fields become highly inhomogeneous and alternating in sign at scale of less than 50 μm. Unlike ceramic the films have critical temperature of trapping lower than the upper temperature of magnetoresistance disappearance. The experimental results are explained by a model in which the magnetic fields are trapped in superconducting loops embedded in Josephson weak links medium. The loops nature which is essentially different for films and ceramics is discussed. Observed temperature and field dependences of trapped field are in good agreement with those calculated for normal law of the loops distribution on critical fields.

## 1. Introduction

Magnetic flux trapping (MFT) in a resistive transition region of granular HTSC is not studied enough up to now. The main reason of this is that in this case the magnetic fields trapped in superconducting regions are closed through normal regions of the HTSC. As a result the arising trapped magnetic fields (TMF) must be sign-varying and can be unobserved by Hall probe and magnetometer SQUID .

Fortunately, in granular superconductors these sign-varying TMF destroy the superconductivity of weak links network and so give rise to frozen magnetoresistance (FMR) [1] and the latter can be used for MFT investigations in the region of resistive transition of granular HTSC [2].

Below we present the results of comparative study of the temperature and field dependences of FMR and TMF in (Pb)Bi-ceramics and Bi-based magnetron-sputtered films near superconducting transition region.

## 2. Samples and measuring techniques.

Two HTSC systems with extended resistive transition were studied: ceramics of nominal composition $Pb_{0.5}Bi_2Sr_2Ca_2Cu_3O_{11}$ with upper transition temperature $T_c$ = 108-110 K and width of transition $\Delta T_c$ = 10-15 K and magnetron-sputtered films of nominal composition $Bi_2Sr_2Ca_2Cu_3O_{10}$ with $T_c$ = 80-82 K and $\Delta T_c$ = 20-40 K.

The ceramics samples dimensions were 3 * 4 * 8 mm$^3$, films dimensions - 3 μm * 4 mm * 8 mm.

The MFT was realized by applying the magnetic fields of magnitude $H_i$ = 1-200 Oe. In ZFC regime the field pulses duration was 30 sec. Measurements of magnetic field and mean value of TMF were made by Hall probe with sensitivity area of 150 * 450 μm$^2$. The threshold sensitivity and experimental error of Hall apparatus were 0.1 Oe. The resistivity measurements were made by the four-point method with device sensitivity of $10^{-5}$ Ω.

## 3. Results and Discussion

The resistive transition width in studied samples essentially increases (by 20-30 K) as a result of MFT or application of the magnetic field H = 20-200 Oe. Moreover the FMR reaches the values equal to a magnetoresistance caused in ceramics by field H = 20 Oe at T = 77.4 K and in films H = 100 Oe at T = 10 K ( Fig. 1), while the averages of TMF measured by Hall probe are negligible (< 0.1 Oe).

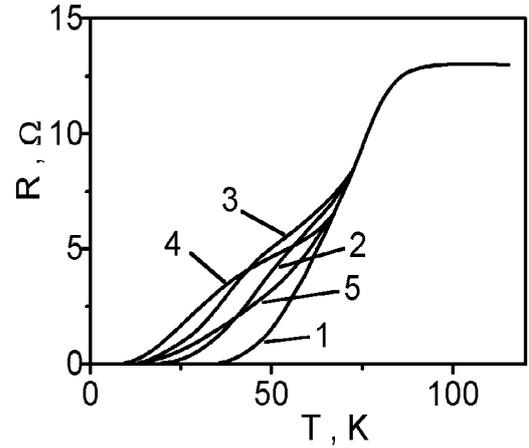

*Fig. 1. Temperature dependences of the Bi-HTSC film resistance. 1 - resistance in zero field, 2 and 3 - magnetoresistance for H = 40 Oe and H = 95 Oe, 4 and 5 - FMR for $H_i$ = 200 Oe at heating from 5 K in FC and ZFC regimes.*

It means that the local TMF which determine the FMR in the granular systems are really rather high (~100 Oe) and sign-varying at scale of less than 50 μm.

The sign-varying TMF can be characterized by an effective value $H_{teff}$ equal to an external field that causes the same magnetoresistance as the trapped fields do. Using this definition the dependences $H_{teff}(H_i,T)$ were determined from FMR and magnetoresistance measurements (Fig. 2).

The other typical feature of films FMR and magnetoresistance dependences on temperature is that the FMR and hence TMF disappear at temperature $T_{tc}$ which is lower by ~10 K than the temperature of magnetoresistance disappearance $T_{mc}$ (Fig. 1). In ceramics the temperatures of FMR and magnetoresistance disappearance coincide, however critical temperature of trapping $T_{tc}$ as determined from $H_{teff}(T)$ turns out to be by 3-5 K higher than $T_{mc}$ .

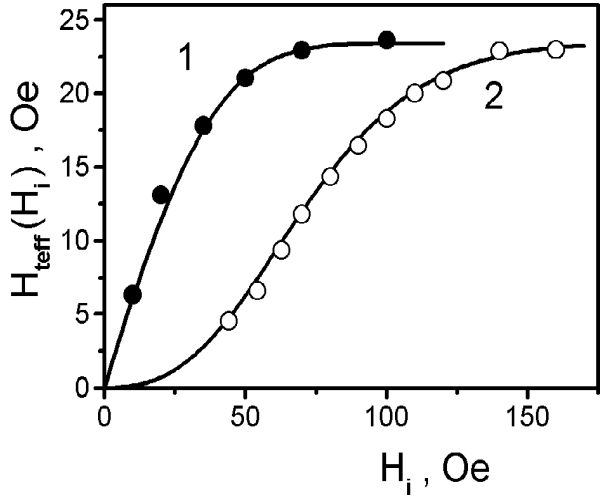

*Fig. 2. Dependence of effective trapped field on inducing pulse field for ceramics, $H_{teff}(H_i)$. 1- FC and 2- ZFC regimes. Solid line - calculated curves.*

One can give the following picture of the change in the state of a granular film as the temperature is lowered. At $T_{mc} < T < T_c$ the individual granules undergo transition to the superconducting state and the film resistance begins to decrease, while magnetoresistance and magnetic flux trapping are insignificant. At $T < T_{mc}$ the weak links that are very sensitive to magnetic fields arise between granules. Besides the additional drop in resistance this gives rise to magnetoresistance. As the temperature decreases lower than $T_{tc}$ the closed superconducting loops are formed. In these loops the MFT occurs, resulting in the frozen magnetoresistance due to the destruction of the weak links of superconducting channels.

Unlike the above the MFT in ceramics can occurs in grains as they undergo transition in superconducting state. However, the FMR can appear only when the weak links arise between grains. We suppose that in ceramic samples the MFT is realized in closed loops of 2223 phase formed on the surface of grains of 2212 phase [3].

The experiments on remagnetization of samples affirm the loops model, which unlike Bean model predicts equal values of FMR obtained as a result of single and of double remagnitization.

The superconducting loops model [2] can be used for quantitative comparison with experimental data. For example let us consider field dependences of TMF. In the FC regime the MFT occurs in all loops. In loops with critical fields $H_c < H_i$ the TMF are proportional to $H_c$, in loops with $H_c > H_i$ - to $H_i$ and one can write:

$$H_{teff} = \int_0^{H_i} H_c \cdot f(H_c) dH_c + H_i \cdot \int_{H_i}^{\infty} f(H_c) dH_c \quad (1)$$

Here $f(H_c)$ is the loops distribution function on $H_c$.

For ZFC case the second term in (1) is absent.

A reasonable agreement of calculation results with experimental data was obtained for normal distribution of loops on $H_c$. For example, for ceramics good agreement was obtained for mean critical field $H_{cm} = 32$ Oe and standard deviation $\Delta H_c = 24$ Oe at 77.4 K (Fig. 2).

So the loops model gives the proper description of MFT near superconducting transition of granular HTSC.